\documentclass[conference]{IEEEtran}
\usepackage{cite}
\usepackage{amsmath,amssymb,amsfonts}
\usepackage{graphicx}
\usepackage[hidelinks]{hyperref}

\begin{document}

\title{Silent Metronome: Rhythmic Grounding \\ for Live Music Accompaniment}

\author{\IEEEauthorblockN{Kevin Bretz, Derya Soydaner, and Aske Plaat}
\IEEEauthorblockA{\textit{Leiden Institute of Advanced Computer Science (LIACS), Leiden University}, Leiden, The Netherlands \\
k.o.bretz@umail.leidenuniv.nl, d.soydaner@liacs.leidenuniv.nl, a.plaat@liacs.leidenuniv.nl}}

\maketitle

\begin{abstract}
Live accompaniment models generate music for an incoming audio stream, committing to each output frame before hearing what comes next. In this strictly causal setting the model must infer tempo, meter, and metrical phase from its own imperfect past, whereby compounding errors quickly become audible as rhythmic drift. Put simply, the model has ears but no temporal reference, so when the ears hear imperfect, ambiguous music, the model will produce a flawed output. We propose Silent Metronome (SiMe), which gives it the temporal reference, encoding the phase within the beat and within the bar as periodic functions, pairing them with tempo and time signature, and supplying the result as a separate conditioning channel. Because this reference is independent of the generated audio, it cannot drift. Complementary auxiliary heads shape the latent representation, including a novel head that predicts the model's own future tokens. With the metrical signal taken from ground-truth annotations, beat alignment improves by a factor of 3.2 over the strictly causal baseline and surpasses a non-causal reference granted a full second of look-ahead. Coherence between input and accompaniment stays within a single point of that reference. These results suggest that streaming accompaniment systems should treat rhythm as a signal to be shared, as human ensembles do, rather than inferred.
\end{abstract}

\begin{IEEEkeywords}
Music Generation, Real-Time Generation, Causal Generation, Human-Computer Interaction (HCI)
\end{IEEEkeywords}

\section{Introduction}
Generative AI has seen growing interest in tools for real-time (online) collaboration and dynamic interaction --- from playable game engines \cite{valevski2025gamengen} and live image synthesis \cite{kodaira2025streamdiffusion} to full-duplex spoken dialogue \cite{defossez2024moshi} and real-time robot manipulation \cite{black2025rtc}. In creative domains, such tools can complement the skills of trained artists much like a collaborative partner. In contrast to prompt-based generation, which compresses the creative process into the lossy realm of language \cite{casacuberta2025disembodied} and strips it of artistic intent \cite{kreminski2026endless}, interactive systems return expressive freedom to the artist's hands.

We focus on a musical application, contributing towards a real-time jamming agent capable of anticipatory accompaniment and expressive improvisation. Recent work has laid the foundation for this trajectory: reinforcement-learned chord accompaniment in the symbolic domain \cite{realchords}, its deployment in a live jamming interface \cite{realjam}, and most recently accompaniment generation directly on streaming audio \cite{wu2025streaming, pasini2026liveband, karchkhadze2026realtime}. However, the strictly causal setting remains stubbornly hard. Each output frame is committed before the input it accompanies has been heard, the model deploys on its own imperfect past after training on flawless ground truth, and this exposure bias decays rhythmic alignment within a few autoregressive inferences. We therefore ask whether the metrical grid must be inferred from the audio stream at all, or whether it can be supplied as an external signal that restores alignment without any look-ahead.

\begin{figure*}[!t]
\centering
\includegraphics[width=0.74\textwidth]{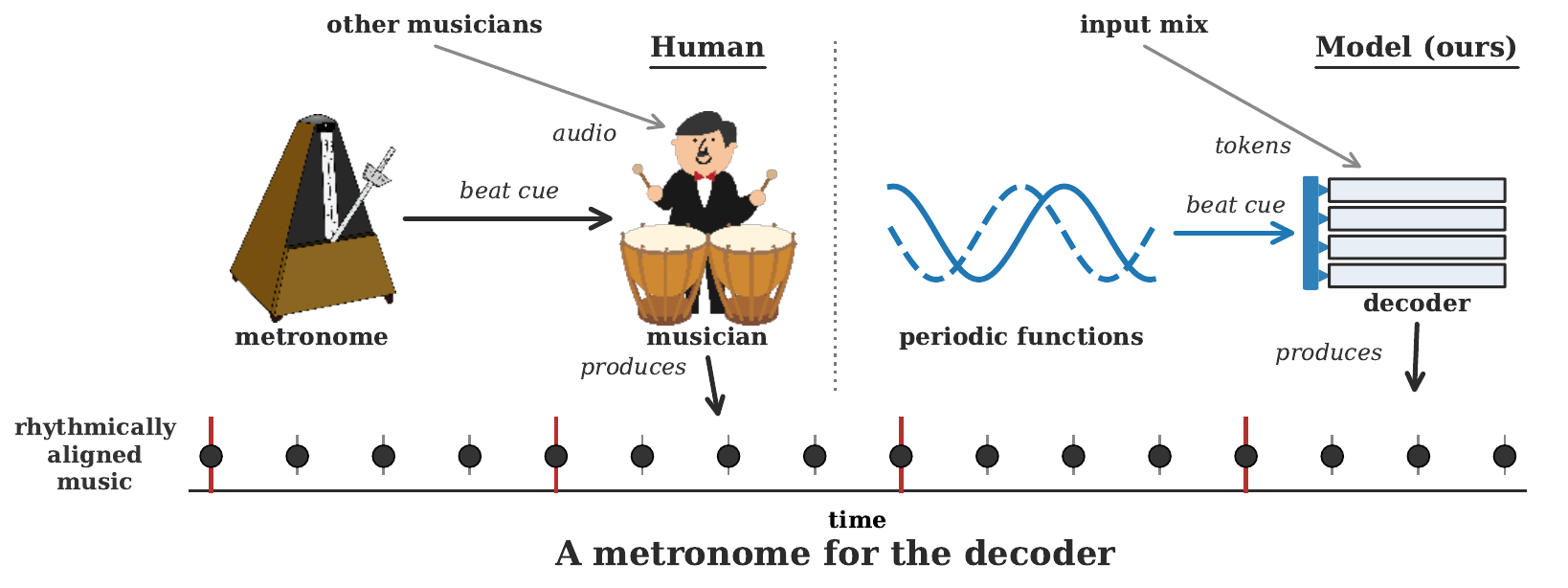}
\caption{The idea behind Silent Metronome. Musicians stay locked onto the metrical grid by following an external beat cue, provided by a metronome, while playing together (left). Our model receives an analogous per-frame periodic signal as a conditioning channel into each decoder block (right). Both produce rhythmically aligned music through an independently sourced temporal signal.}
\label{fig:headline}
\end{figure*}

Metaphorically speaking, the model has ears but no sense of time. Silent Metronome (SiMe) hands it a watch, in the form of an external metrical signal provided through a separate input channel (Fig.~\ref{fig:headline}) that decouples rhythm from the causal audio input. During deployment the signal can be clock-given, from a Digital Audio Workstation (DAW) or shared click track, or, in theory, live-tracked, estimated causally from the incoming audio. All experiments below use the exact dataset grid, an oracle upper bound that isolates the anchoring mechanism from the separate problem of live estimation. To also target harmonic coherence, we layer latent-shaping auxiliary heads on top, including a future token prediction head that is, to the best of our knowledge, new to music generation.

We build directly on the streaming framework of Wu \emph{et al.}~\cite{wu2025streaming}, leaving its architecture, codec, and training recipe unchanged, so any improvements are attributable to the added signal and objectives alone. Concretely, we contribute (i) per-frame metrical-phase conditioning for causal music generation, injected through adaptive layer normalization, (ii) a target-token future auxiliary head that pressures the decoder to plan ahead, and (iii) ablations that provide a definitive answer to our question: using a grid with perfect information, the strictly causal model equipped with both extensions more than triples its beat alignment, surpassing alternatives that use only a single extension, tempo and meter conditioning without phase components, or even a full second of look-ahead.

Our code, checkpoints, audio samples, and extended results are available via our \href{https://kevin-bretz.github.io/projects/silentmetronome/}{project page}.\footnote{\url{https://kevin-bretz.github.io/projects/silentmetronome/}}

\section{Related Work}
For an expansive overview of live music agents we refer to Kim et al.~\cite{kim2026designspace}.

\emph{Live accompaniment.} ReaLchords \cite{realchords} demonstrates learned online chord accompaniment, and ReaLJam \cite{realjam} deploys it in a live jamming interface, but both make use of symbolic data that lives on a fixed grid, trivializing the objective of beat alignment. Wu \emph{et al.}~\cite{wu2025streaming} move accompaniment into the audio domain with a streaming music generation (SMG) framework that we build on directly (Sec.~\ref{sec:background}), observing that a look-ahead buffer greatly benefits overall beat alignment and coherence between the input context sequence and the generated output, whereas removing it or simulating latency makes the model rely mainly on its own generation history, dropping coherence with the input toward the random-pairing floor. LiveBand \cite{pasini2026liveband} improves on the SMG baseline across audio quality, beat alignment, and coherence, and explicitly measures rhythmic drift over time, and Karchkhadze and Dubnov approach the same setting with sliding-window latent diffusion in place of autoregression \cite{karchkhadze2026realtime}. However, scores reported across these systems are not directly comparable, as codecs, metric implementations, and evaluation windows differ, so we compare against our retrained SMG baseline. Crucially, none of these hand the generative model an explicit, periodic reference for \emph{where in the bar it currently is}.

\emph{Exposure bias and autoregressive drift.} Causal autoregressive generation suffers from \emph{autoregressive drift}, where errors accumulate over a rollout because the model conditions on its own imperfect past, an exposure-bias issue prominent across causal generation \cite{valevski2025gamengen}. In music it manifests as tempo and meter changing over time, known as \emph{rhythmic drift}. Most existing solutions are employed during training, such as scheduled sampling \cite{bengio2015scheduled}, noise augmentation on context frames \cite{valevski2025gamengen}, per-token noise schedules \cite{chen2024diffusionforcing}, and training on the model's own rollouts \cite{huang2025selfforcing}. LiveBand \cite{pasini2026liveband} instead removes the feedback loop entirely, so generated latents are never fed back as input, and generation is driven by noise and the input mix under sequence-level adversarial training.

\emph{Conditional generation and metrical structure.} Conditioning channels are standard in offline music generation, such as text and melody in MusicGen \cite{copet2023musicgen}, timing controls in Stable Audio Open \cite{evans2025stableaudioopen}, and stem-aware accompaniment in MusicGen-Stem \cite{rouard2025musicgenstem} and STAGE \cite{strano2025stage}. Music ControlNet \cite{wu2024musiccontrolnet} demonstrates time-varying rhythm controls for offline diffusion models. Explicit cyclic-phase conditioning also appears wherever behavior locks to a periodic grid, in character controllers driven by gait phase \cite{holden2017pfnn}, locomotion policies fed sinusoidal clocks \cite{siekmann2021bipedal}, and vocoders excited by pitch-locked periodic functions \cite{wang2019nsf}. Beat trackers such as Beat This! \cite{foscarin2024beatthis} and madmom \cite{bock2016madmom} recover beat grids reliably enough to serve as evaluation tools and as candidate sources of our conditioning signal, the former offline for precomputed material, the latter causally for live use. Our work adapts periodic metrical conditioning to the strictly causal streaming regime, where it acts as an anchor against drift that can be added on top of any of the previously mentioned alternatives, injecting it through the adaptive normalization of diffusion transformers~\cite{peebles2023dit} but modulating per frame rather than once per sample.

\section{Method}

\subsection{Background}
\label{sec:background}
For our experiments we build on the streaming accompaniment framework of Wu \emph{et al.}~\cite{wu2025streaming}, a prefix-decoder transformer that takes a live audio stream as input and autoregressively generates tokens for the accompanying stream. It operates on tokens from a causal variant of the Descript Audio Codec (DAC)~\cite{kumar2023dac}, where encoder and decoder are both streamable, at a 50\,Hz frame rate with four residual codebooks per frame arranged via the delay pattern of Copet \emph{et al.}~\cite{copet2023musicgen}. The design is centered on two scalars: \emph{future visibility} $f_v$ defines the gap in frames between the latest input frame the model sees and the current output frame, used to simulate latency ($f_v<0$) or a look-ahead buffer ($f_v>0$), and the \emph{chunk size} $k$ defines the number of frames generated per inference. Throughout this work we maintain $k=50$ (one-second chunks), matching the chunk length used by the baseline, and we prioritize the strictly causal application of $f_v=0$ frames, with a single non-causal reference at $f_v=+50$ for comparison.

\subsection{The Silent Metronome}
\label{sec:cond}

\begin{figure}[!t]
\centering
\includegraphics[width=\columnwidth]{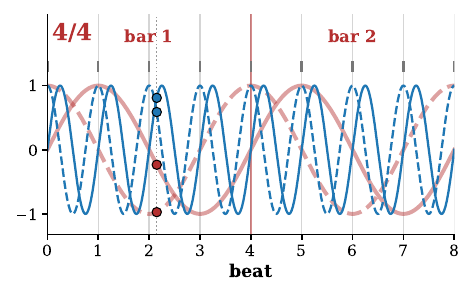}
\caption{The four beat-phase conditioning channels over a two-bar $4/4$ example, sine (solid) and cosine (dashed) of the beat phase (blue) and of the bar phase (red). Positive-gradient zero-crossings of beat phase sit on beat boundaries, the bar phase completes one cycle per bar, and the dotted line picks out the four values carried by a single frame.}
\label{fig:signals}
\end{figure}

Musical \emph{meter} organizes a song as a periodic structure, a steady sequence of \emph{beats} grouped into recurring \emph{bars} by the time signature, so every timepoint of a performance has a well-defined position on the resulting grid. Preprocessing resolves the exact tempo map and time signature of every track (Sec.~\ref{sec:data}) to beat onsets, bar positions, tempo, and time signature at the 50\,Hz token rate, while note timing follows the source MIDI files, largely but not universally quantized to this grid. For a frame $t$ inside a beat that starts at frame $b_i$ and spans $\ell_i$ frames, the \emph{beat phase} is $\phi_\text{beat}(t) = \left((t-b_i)/\ell_i\right) \bmod 1$. With $n$ beats per bar and $\beta_i \in \{0,\dots,n{-}1\}$ the bar position of beat $i$, the \emph{bar phase} is $\phi_\text{bar}(t) = \left((\beta_i + \phi_\text{beat}(t))/n\right) \bmod 1$. Each frame then carries
\begin{equation}
\boldsymbol{\phi}_t = \big[\sin 2\pi\phi_\text{beat},\, \cos 2\pi\phi_\text{beat},\, \sin 2\pi\phi_\text{bar},\, \cos 2\pi\phi_\text{bar}\big],
\label{eq:beatcond}
\end{equation}
a wrap-free encoding whose sine/cosine pairing makes the position on the beat and bar circles unambiguous and continuous across boundaries (Fig.~\ref{fig:signals}). The full conditioning vector extends the phase quadruple with explicit tempo and meter descriptors,
\begin{equation}
\mathbf{c}_t = \mathrm{MLP}\big(\big[\boldsymbol{\phi}_t;\; \tau_t;\; \bar{\tau};\; \delta_\text{tempo};\; \delta_\text{meter};\; \mathbf{e}_\text{num};\; \mathbf{e}_\text{den}\big]\big) \in \mathbb{R}^{d},
\label{eq:condvec}
\end{equation}
where $\tau_t = \log(\mathrm{BPM}_t/120)$ is the per-frame local log-tempo, $\bar{\tau} = \log(\overline{\mathrm{BPM}}/120)$ its window-level counterpart, $\delta_\text{tempo}, \delta_\text{meter} \in \{0,1\}$ flag tempo or meter changes within the window, and $\mathbf{e}_\text{num}, \mathbf{e}_\text{den} \in \mathbb{R}^{8}$ are learned embeddings of the time-signature numerator and denominator. This results in a total of 24 input channels, which are projected by a two-layer multilayer perceptron (MLP) to the model width $d$. At training and evaluation time the metrical grid comes directly from the dataset's symbolic representation.

We inject the metrical signal into the decoder via the AdaLN-Zero mechanism of Peebles and Xie~\cite{peebles2023dit}. A shared MLP expands $\mathbf{c}_t$ by a factor $m$, and within every transformer block, zero-initialized linear maps predict a multiplicative gain for each LayerNorm together with a sigmoid-gated residual scale on each branch, so that every block starts as the identity function and learns how strongly to listen to the meter (our implementation predicts gains and gates, without additive shifts). Per-frame modulation lets every layer specialize to the current phase, and because the metrical signal is independent of past audio outputs, rhythmic conditioning is immune to autoregressive drift.

The conditioning channel adds $278$M parameters ($497$M conditioned vs.\ $219$M unconditioned), but since the modulations depend only on the externally supplied grid, known one chunk in advance, they are precomputed for a whole chunk in parallel, leaving attention and KV-cache costs untouched. Measured end-to-end on a single A100, conditioning adds roughly $5\,\%$ to per-chunk generation time. Capacity alone does not explain the gains, as the phase-stripped variant in Table~\ref{tab:headline} is parameter-identical, its phase channels merely held at zero, yet stays at the baseline's alignment.

\subsection{Latent Shaping and Target-Token Future Auxiliary Heads}
\label{sec:aux}
To address not only beat alignment but also harmonic and textural expression, we attach auxiliary heads to the hidden state, trained jointly and dropped at inference. Two shallow MLP heads predict the target stem's \emph{current} content, consisting of active pitches (128 binary values, known as multi-pitch) and an 84-bin constant-Q spectrum~\cite{brown1991cqt}. Such pitch- and spectrum-prediction objectives are established tools for instilling musical structure in audio networks~\cite{manilow2020cerberus, li2024mert}. The third head brings multi-token prediction to music generation, to the best of our knowledge for the first time. In language modeling, predicting multiple future tokens is known to encourage planning beyond the immediate next token~\cite{qi2020prophetnet, gloeckle2024mtp}. A two-layer MLP predicts the output DAC tokens at positions $p+\delta$ for $\delta \in \{10, 25, 40\}$ frames (0.2, 0.5, and 0.8\,s ahead), with a per-codebook cross-entropy (CE) averaged over valid positions. This pressures the internal representation to also function predictively. The total objective is
\begin{equation}
\mathcal{L} \;=\; \mathcal{L}_\text{token} + \lambda_\text{mp}\,\mathcal{L}_\text{mp} + \lambda_\text{cqt}\,\mathcal{L}_\text{cqt} + \lambda_\text{fut} \sum_{\delta}\mathcal{L}_\text{CE}^{(\delta)},
\label{eq:loss}
\end{equation}
with all weights set to $1$.

\section{Experiments}

\subsection{Dataset}
\label{sec:data}
All experiments use Slakh2100~\cite{manilow2019slakh}, 2{,}100 multitrack songs synthesized from MIDI with professional virtual instruments, with the SMG baseline's train, validation, and test splits~\cite{wu2025streaming}. Because the audio is rendered from symbolic data, the tempo map and time signature of every track are known exactly, providing the noise-free metrical grid this study requires, and the dataset remains the common benchmark of causal accompaniment work~\cite{wu2025streaming, pasini2026liveband, karchkhadze2026realtime}. Real-audio recordings offer no such grid, so evaluation beyond synthesized data requires tracker-derived annotations, which we defer to the live-tracked setting. Data consists of ten-second windows (500 frames at 50\,Hz), with one stem held out as the generation target and the remaining stems mixed down as the input.

\subsection{Setup and Evaluation Protocols}
We focus on training using perfect conditioning information, identifying the upper bound our methods can produce when such information is available, and leaving the non-trivial task of causally predicting said conditioning signal for a fully self-contained system to future work. Every variant trains for 200k steps at batch size 16 with AdamW (peak learning rate $10^{-4}$, 10k-step warmup, cosine decay to $10^{-5}$) in bf16 precision on a single A100. Evaluation follows the baseline protocol, with 1{,}024 held-out windows, continued chunk-by-chunk at $k=50$ with $f_v=0$ unless stated otherwise, and all conditioned variants receive the oracle dataset grid. Each variant is evaluated five times with different sampling seeds, and we report the mean and standard deviation across those runs. \emph{Beat-F} is the standard beat-tracking metric (madmom's \texttt{BeatEvaluation}~\cite{bock2016madmom}), computed between beats extracted using Beat This!~\cite{foscarin2024beatthis} on the generated accompaniment and on the input mix. It determines whether the accompaniment and the incoming stream agree on the pulse, not whether either fits a static grid. \emph{CoCoLa}~\cite{ciranni2025cocola} scores harmonic and percussive coherence between accompaniment and input, and \emph{FAD}~\cite{kilgour2019fad} (Fr\'echet audio distance, VGGish backbone) measures distributional audio quality. Beyond this headline protocol we score each target class separately, with the held-out ground-truth stem evaluated identically as a reference. To probe generation past the training horizon we adopt the long-horizon drift protocol of LiveBand~\cite{pasini2026liveband}, generating twenty-second continuations with a sliding window at the same one-second chunk size and scoring the two ten-second halves separately.

\subsection{Results and Ablations}

\begin{table}[!t]
\caption{Headline results on the Slakh2100 test split (1{,}024 samples), reported as mean $\pm$ standard deviation over five sampling seeds. Best generated result per column in bold. The ground-truth row scores the held-out real stem under the identical protocol. $f_v$ is the future-visibility offset in frames ($0$ = strictly causal; $+50$ = one chunk of look-ahead, non-causal reference).}
\label{tab:headline}
\centering
\resizebox{\columnwidth}{!}{%
\begin{tabular}{@{}lcccc@{}}
\hline
Method & $f_v$ & Beat-F $\uparrow$ & CoCoLa $\uparrow$ & FAD $\downarrow$ \\
\hline
Baseline                     &  $0$  & $0.133 \pm 0.003$ & $58.56 \pm 0.14$ & $5.56 \pm 0.10$ \\
+ Aux only                   &  $0$  & $0.133 \pm 0.002$ & $57.93 \pm 0.09$ & $5.46 \pm 0.10$ \\
+ Cond (tempo + time sig.\ only) &  $0$  & $0.141 \pm 0.006$ & $57.13 \pm 0.08$ & $4.80 \pm 0.04$ \\
+ SiMe (full cond)           &  $0$  & $0.380 \pm 0.003$ & $60.03 \pm 0.04$ & $4.25 \pm 0.06$ \\
+ SiMe + Aux (pitch, spectrum) &  $0$  & $0.411 \pm 0.011$ & $60.36 \pm 0.07$ & $\mathbf{3.96} \pm 0.05$ \\
+ SiMe + Aux + future head   &  $0$  & $\mathbf{0.432} \pm 0.008$ & $60.84 \pm 0.05$ & $4.38 \pm 0.10$ \\
\hline
Non-causal ref.              & $+50$ & $0.269 \pm 0.006$ & $\mathbf{61.70} \pm 0.05$ & $5.29 \pm 0.08$ \\
Ground truth                 & --    & $0.570$ & $66.27$ & -- \\
\hline
\end{tabular}%
}
\end{table}

Metrical conditioning alone lifts Beat-F from $0.133$ to $0.380$ (Table~\ref{tab:headline}). Adding the auxiliary heads brings it to $0.432$, a factor of $3.2$ over the causal baseline, well beyond the non-causal reference at $f_v{=}{+}50$ ($0.269$), which is granted a full second of look-ahead, and about three quarters of the ground-truth's $0.570$. The refinement decomposes across the last two rows, as the pitch and spectrum heads carry alignment to $0.411$ and the future-token head contributes the remaining step. CoCoLa rises alongside, from $58.56$ to $60.03$ before any auxiliary heads. CoCoLa itself rewards shared rhythmic placement, but its harmonic component rises as well, so the gain is not confined to the rhythm channel. With the auxiliary heads on top, the CoCoLa score climbs to $60.84$, within a single point of the non-causal $61.70$. FAD improves as well, from $5.56$ for the baseline to $4.25$ with SiMe alone, and a best of $3.96$ with the pitch and spectrum heads added. Layering the future-token head on top of everything else trades some distributional fidelity, now at $4.38$, for further alignment and coherence gains. The auxiliary heads alone leave Beat-F at $0.133$, indistinguishable from the baseline, and the conditioning stripped of its phase channels reaches only $0.141$, so the metrical phase is the core mechanism and the auxiliary heads merely refine what it provides.

Per class, the headline model reaches $0.729 \pm 0.012$ Beat-F on drum targets against the ground-truth stems' $0.777$, $0.571 \pm 0.027$ on bass targets against $0.722$, and $0.363 \pm 0.010$ on harmonic targets against $0.511$, reaching about 94\%, 79\%, and 71\% of the ground-truth scores, and its advantage over the non-causal reference ($0.373$, $0.391$, and $0.230$) holds everywhere. Under the drift protocol, where the harder second half costs even the ground-truth reference $4\,\%$ of its score, the conditioned models track that content floor, with relative Beat-F declines of $3\,\%$ for SiMe and $5\,\%$ for the full model, maintaining $75\,\%$ of the ground-truth score in both halves, while the unconditioned rows lose $14$--$16\,\%$ and the non-causal reference loses $13\,\%$, three to four times the floor. The external signal carries alignment past the training horizon which a second of look-ahead does not.

\subsection{Discussion and Future Work}
Three observations are worth noting. \emph{First}, the metronome agreeing with the input is the premise of a synchronized clock, and is inherently given when using MIDI-generated data, yet it leaves untested how the model responds when a live partner drifts from the set rhythm. A more robust model should be trained and tested on real-audio recordings, where the input is not fundamentally aligned to the metrical grid.
\emph{Second}, per-instrument recovery matches our informal listening, wherein percussive accompaniment profits most from the purely metrical signal while harmonic targets leave the largest share unrecovered. To address the harmonic dimension equally, a harmonic counterpart to the silent metronome, such as a chord chart channel, is a reasonable extension. \emph{Third}, our models receive a signal the references do not. However, the signal is cheap, external, and drift-free, ideal for streaming systems with access to a shared clock, the norm in DAW, studio, and stage setups. Without one the grid must be estimated causally from the incoming audio, so evaluating and retraining under tracker-derived conditioning, alongside the delayed-observation regime ($f_v<0$) where the model must compensate real-world latency, is the natural next step.

\section{Conclusion}
We introduced Silent Metronome, a periodic encoding of metrical position injected per frame into every decoder block of a streaming accompaniment model, complemented by an auxiliary head that predicts the model's own future tokens. Because the signal is independent of the generated audio it cannot drift, and it costs a live setting nothing beyond the click track it already has. Under strictly causal operation beat alignment triples, surpassing a non-causal reference granted a full second of look-ahead, and the advantage persists past the training horizon. Rhythm, it turns out, need not be inferred from the audio stream at all, it can simply be told to the model, the way a metronome tells a band.

\IEEEtriggeratref{19}

\end{document}